\documentclass[11pt,a4paper]{article}

\usepackage[normalem]{ulem}
\usepackage[utf8]{inputenc}
\usepackage{amssymb, amsmath, amsbsy} 
\usepackage{authblk}
\usepackage{float}
\usepackage{graphicx}   
\usepackage{hyperref}
\usepackage{subfigure}

\title{Geometric Construction of non-linear Sigma models \\ with $Q$-ball/$Q$-kink solutions}

\author[1,2]{A. Alonso-Izquierdo}
\author[2]{ D. Canillas Martínez}
\author[2]{C. Garzón Sánchez}
\author[1,2]{M.A. González León}

\affil[1]{ Departamento de Matemática Aplicada, University of Salamanca,
Casas del Parque 2, 37008 - Salamanca, Spain.}
\affil[2]{ IUFFyM, University of Salamanca,
Plaza de la Merced 1, 37008 - Salamanca, Spain}
\date{}

\begin{document}

\maketitle

%%%%%% Abstract %%%%%%
\begin{abstract}
Non-linear Sigma models involving $U(1)$ symmetry group are studied using a geometrical formalism. In this type of models, $Q$-balls and $Q$-Kinks solutions are found. The geometrical framework described in this article allows the identification of the necessary conditions on the metric and the potential to guarantee the existence of these $Q$-balls and $Q$-Kinks. Using this procedure, Sigma models where both types of solutions coexist, have been identified. Only the internal rotational frequency distinguishes which one of these defects will arise.
\end{abstract}

%PACS: 11.15.Kc; 11.27.+d; 11.10.Gh

%%%%%%%%%%%%%%%%%%%%%%%
%%%%%% Main Text %%%%%%
%%%%%%%%%%%%%%%%%%%%%%%

\section{Introduction}

Scalar field theories with a global $U(1)$ symmetry can present time-dependent non-topological solitons known as $Q$-balls. This kind of solutions conserves a Noether charge associated with the $U(1)$ symmetry \cite{LEE1992251,shnir_2018}. The time dependence of these solutions allows them to avoid the severe restrictions of Derrick's theorem \cite{derrick1964comments,manton2004topological}, so they can arise in theories with several spatial dimensions. These non-topological solutions were first introduced by Friedberg, Lee, and Sirling \cite{friedberg1976class} in 1976. The authors studied a theoretical model involving a complex scalar field coupled to a real scalar field in three space dimensions. Later, Coleman revisited the study of $Q$-balls in systems with one complex scalar field \cite{coleman1985q, coleman1986errata}, investigating their properties. 
These solutions take the form $\phi=f(r) e^{i \omega t}$ where $\phi$ is the complex field, and $f(r)$ is a monotonically decreasing function of the distance from the origin that goes to zero when $r$ tends to infinity. The field rotates in the internal space with a constant frequency $\omega$. It can be demonstrated that depending on the chosen potential, this frequency must be bounded as $\omega_{-}^2<\omega^2<\omega_{+}^2$ to guarantee the existence of $Q$-ball type solution.

Among all the applications associated to these objects, $Q$-balls have garnered substantial attention because it was suggested that such solutions could play an important role in diverse scenarios of the evolution of the early universe. The most compelling physical interest comes from the fact that $Q$-balls can be produced in the early stages of the universe in supersymmetric extensions of the Standard Model, where one finds leptonic and baryonic $Q$-balls related to conservation of lepton and baryon numbers respectively \cite{KUSENKO1997108}. In addition, it has been argued that these solutions may play an important role in the baryogenesis by means of the Affleck-Dine mechanism \cite{Dine2003,AFFLECK1985361}. They were also considered a plausible candidate for Dark Matter \cite{KUSENKO199846}. Finding models that present $Q$-balls in which analytical solutions exist is quite challenging. However, once the analytical expression of these solitons have been found, it is possible to study the properties of these objects in depth. Some scenarios have been considered with the aim to obtain analytical solutions in theories with one complex scalar field in $(1+1)$ dimensions \cite{bazeia2016exact,bazeia2016compact,bazeia2019quasi,bazeia2017split,Alonso2023_1}. In these articles, the authors focus mainly on attaining analytical solutions with different features, like models involving compact $Q$-balls.

$Q$-kinks are a similar type of solutions which carry a topological charge in addition to the Noether charge  \cite{LEESE1991283,Alonso2023_2,Loginov:2011zz,lensky2001domain}. In fact, these particular objects can be interpreted as topological kinks that rotate in the internal space. This idea was developed in \cite{Loginov:2011zz}, where Loginov showed that the non-linear $O(3)$ Sigma model admits a non static generalization of the sine-Gordon kink. The first works where $Q$-kinks were found deal with $(4, 4)$-Supersymmetric $(1+1)$-dimension Sigma models \cite{ABRAHAM199285,ABRAHAM1992225}. This specific type of solitonic solutions are present in Sigma models in the context of String theory \cite{HEWSON1997201, Eric_Bergshoeff_1999, LAMBERT2000606, PhysRevD.63.085002}. For example, the application of T-duality transformations of BPS states in massive $(1+1)$ Supersymetric Sigma models are discussed in \cite{HEWSON1997201}, where an explicit example of $Q$-kink is given. In the study of Supersymmetric Sigma models with a Kähler target space, energy bounds are derived for planar and compactified M2-branes. $Q$-kink solitons saturate these bounds, preserving half of supersymmetry \cite{Eric_Bergshoeff_1999}.

In the present article, we develop a geometrical framework that allows the construction of $(1+1)$-dimension non-linear Sigma models, in which can be found $Q$-ball and $Q$-kink solutions. This formalism helps to understand these objects within the same perspective, since the form of the metric of the internal space together with the distribution of the zeros of the potential term, and the internal rotation frequency, will determine whether the solutions in our model are $Q$-balls or $Q$-kinks.

The structure of this paper is outlined in the following manner. In Section 2, we introduce the general formalism of Sigma models which admit $Q$-ball and $Q$-kink solutions in different geometries of the internal space. In Section 3, some examples of these theories in $(1+1)$ dimension are given, on which $Q$-ball and $Q$-kink solutions have been analytically found and which can be used to illustrate the previous theoretical framework. Finally, we conclude the article by detailing the main results obtained. 

\section{General formalism for Sigma models with $Q$-ball/$Q$-kink solutions}

The existence of $Q$-balls and $Q$-kinks in certain theories underlies the presence of a cyclic coordinate in the scalar field theory where such solutions arise. The existence of these variables implies the conservation of a Noether charge. Two simple examples can be used to illustrate this claim. Firstly, notice that the action functional for the standard complex field theory giving rise to $Q$-balls in $(1+1)$-dimensions
\begin{equation}
S_1[\phi,\overline{\phi}]=\int dx \int dt\left[\,\dfrac{1}{2} \,\partial_\mu \overline{\phi} \, \partial^\mu \phi - V\left(|\phi|\right)\right] \label{eq:action0}
\end{equation}
can be rewritten after using the representation $\phi(x,t)=\rho(x,t) e^{i\theta(x,t)}$ as
\begin{equation}\label{eq:action1}
S_1[\rho,\theta]=\int dx \int dt\left[\dfrac{1}{2} \partial_\mu \rho \, \partial^\mu \rho+ \dfrac{\rho^2}{2} \partial_\mu \theta \, \partial^\mu \theta -  V\left(\rho \right)\right]\ .
\end{equation}
The Minkowski metric $\eta_{\mu \nu}$ is chosen as $\eta_{00}=-\eta_{11}=1$ and $\eta_{01}=\eta_{10}=0$, the Einstein summation convention is used both in Latin and Greek indices and the notation $x^0=t$, $x^1=x$ is employed. From (\ref{eq:action1}) it is clear that the phase $\theta(x,t)$ of the complex field $\phi(x,t)$ is a cyclic coordinate. Secondly, the non-linear $O(3)$ Sigma model addressed by Loginov in \cite{Loginov:2011zz} reads
\begin{equation}
S_2[\phi^1,\phi^2,\phi^3]=\int dx \int dt\left[\dfrac{1}{2} \, \partial_\mu \phi^i \, \partial^\mu \phi^i - \lambda^2 \left(1-\phi^3 \right)\right] , \hspace{0.5cm} i=1,2,3 \label{eq:action2}
\end{equation}
where $\phi^i \phi^i= 1$ constrains the fields to take values on the sphere $\mathbb{S}^2$ and $\lambda$ is a constant that specifies the breaking of the $O(3)$ symmetry in this theory.  In  \cite{Loginov:2011zz} the $Q$-kinks arising in this model have been analytically identified. By using spherical coordinates in the isotopic (internal) space $\phi^1 =\sin \theta \cos \varphi$, $\phi^2 = \sin \theta \sin \varphi$ and $\phi^3 = \cos \theta$, the action (\ref{eq:action2}) can be expressed as
\begin{equation}
    S_2 [\theta,\varphi]=\int dx \int dt\left[\dfrac{1}{2} \partial^\mu \theta \, \partial_\mu \theta+ \dfrac{1}{2} \,\sin^2 \theta \,\partial^\mu \varphi \, \partial_\mu \varphi - (1-\cos \theta)\right] \label{eq:action3}
\end{equation}
where now the cyclic coordinate is the azimuthal angle $\varphi$. The $Q$-charges in these theories are expressed as the integral over the space of a metric factor multiplied by the time derivative of the cyclic coordinates:
\begin{equation}\label{eq:ChargesExample}
    Q_1=\int_{-\infty}^\infty dx \, \rho^2 \, \partial_0 \theta   \hspace{0.5cm}, \hspace{0.5cm}
    Q_2 =\int_{-\infty}^\infty dx \, \sin^2 \theta \, \partial_0 \varphi \ .
\end{equation}
Taking into account the previously described scenario, the goal of this Section is to construct the general Sigma model where these $Q$-ball/$Q$-kink type solutions can be found. For the sake of concreteness, we shall deal with $(1+1)$-dimensional scalar field theory models\footnote{The generalization to the $(1+d)$-dimensional case is straightforward.} whose dynamics are governed by the action 
  \begin{equation} \label{eq:action}
   S[\phi]=\int dx \int dt\left[\dfrac{1}{2} \eta^{\mu \nu} g_{i j}\left( \phi \right) \partial_\mu \phi^i \partial_\nu \phi^j-V\left(\phi\right)\right] \ . 
  \end{equation}
where $\phi^{i}\in {\rm Maps}(\mathbb{R}^{1,1},\mathbb{R})$ with $i=1,\dots,n$, i.e., the internal space corresponds to a $n$-dimensional space, where $\phi=(\phi^1,\dots,\phi^n)$ is a system of local coordinates, with a Riemannian metric $g_{ij}(\phi)$, and $V(\phi)$ is a potential function that is assumed to be positive semi-definite.

Let us consider that one of the components of the field $\phi$ is cyclic, for example $\phi^n$, without loss of generality, that is, $\frac{\partial \mathcal{L}}{\partial \phi^{n}}=0$. This assumption implies that the potential term $V$ in (\ref{eq:action}) must depend only on the $n-1$ first field components of $\phi$. The same fact is applicable to the metric components $g_{ij}$. In order to alleviate the notation we write the previous conditions as
\[
V =  V(\widetilde{\phi}), \hspace{0.2cm} g_{ij} = g_{ij}(\widetilde{\phi}) ,\hspace{0.5cm} \text{where }\widetilde{\phi}=\left(\phi^{1},\dots, \phi^{n-1} \right) \ .
\] 
The presence of this cyclic coordinate implies the existence of a conserved Noether charge
\begin{equation}\label{eq:QChargeDef}
Q= \int_{-\infty}^\infty g_{n i} 
(\widetilde{\phi}) \ \partial^0\phi^idx \ .
\end{equation}
$Q$-balls/$Q$-kinks are solutions of the field equations
\begin{equation}
\eta^{\mu\nu} \partial_{\mu} \partial_{\nu} \phi^i +\eta^{\mu\nu}\Gamma^{i}_{jk} \partial_\mu \phi^j\partial_\nu \phi^k +g^{ia}\dfrac{\partial V}{\partial \phi^a}=0\label{pde}
\end{equation}
whose total energy 
\begin{equation} \label{energy00}
E\left[\phi\right]=\displaystyle \int d x\left[\dfrac{1}{2}g_{ij}(\widetilde{\phi}) \dfrac{\partial \phi^i}{\partial t}\dfrac{\partial \phi^j}{\partial t}+\dfrac{1}{2}g_{ij}(\widetilde{\phi}) \dfrac{\partial \phi^i}{\partial x}\dfrac{\partial \phi^j}{\partial x} + V(\widetilde{\phi}) \right] 
\end{equation}
is finite. In (\ref{pde}), $\Gamma_{j k}^{i}=\dfrac{1}{2} g^{i m}\left(\partial_{j} g_{k m}+\partial_{k} g_{jm}-\partial_{m} g_{j k}\right)$ are the Christoffel symbols associated to the metric $g_{ij}$. In addition, $Q$-ball/$Q$-kink type solutions spin with a constant internal rotation frequency $\omega$ with respect to the (angular) $\phi^n$-cyclic coordinate while the rest of variables are time independent. Therefore, it is natural to write
\begin{equation}
\phi^a=\phi^a(x) \hspace{0.5cm}, \hspace{0.5cm} \phi^{n}=\omega t\hspace{0.5cm}, \hspace{0.5cm}  \forall a =1,\dots,n-1\, \ , \label{ansatz}
\end{equation}
that reduces (\ref{energy00}) to,
\begin{equation}\label{energy01}
E\left[\phi\right]=\displaystyle \int d x\left[ \dfrac{1}{2}g_{ab}(\widetilde{\phi}) \dfrac{d \phi^a}{d x}\dfrac{d \phi^b}{d x} + \dfrac{1}{2} \omega^2 \, g_{nn}(\widetilde{\phi}) + V(\widetilde{\phi}) \right] 
\end{equation}
and (\ref{eq:QChargeDef}) to
\begin{equation} \label{eq:Q_2dim}
Q= \omega \, \int_{-\infty}^\infty g_{nn}(\widetilde{\phi})\,d x \ .
\end{equation}
Note that the $Q$-charge in (\ref{eq:Q_2dim}) is determined by both the internal rotational frequency and the global metric behavior. The charges (\ref{eq:ChargesExample}) associated to the previously mentioned models are particular cases of this general expression. The finiteness requirement for energy (\ref{energy01}), and the condition of positive semi-definiteness for $V$ lead to the asymptotic conditions
\begin{equation}\label{eq:boundary}
\lim_{x \to \pm \infty}\phi = v_\pm  \in \mathcal{M} \hspace{0.5cm}, \hspace{0.5cm} \lim_{x \to \pm \infty}  g_{ab}(\widetilde{\phi}) \dfrac{d \phi^a}{d x}\dfrac{d \phi^b}{d x}=0 
\end{equation}
where $\mathcal{M}$ stands for the set of vacuum solutions $\mathcal{M}=\{\phi \in \mathbb{R}^{n}:V\left(\phi\right)=0\}$, while in order to keep finite the kinetic contribution in (\ref{energy01}) the condition
\begin{equation}\label{eq:boundary2}
\lim_{x \to \pm \infty}  g_{nn}(\widetilde{\phi}) =  g_{nn}(v_\pm) =0 
\end{equation}
must hold. In the standard $Q$-ball theory described by the action $S_1$ written as (\ref{eq:action1}) in the polar representation of the complex field, $g_{nn} = \rho^2$, which implies that $Q$-balls must be non-topological solitons connecting a vacuum point located at the origin of the internal complex plane with itself. This requirement can be broken, for example, in the case that the internal space consists of a homogeneous space with respect to the rotation group associated to the cyclic coordinates, which could act as a stationary group for some elements of the set of vacua ${\cal M}$. This is the case, for example, when the internal space is given by the sphere $\mathbb{S}^2$ \cite{Loginov:2011zz} where $g_{nn}=\sin^2 \theta$ and thus topological solitons can arise joining two different vacuum points $v_\pm$ characterized by the values $\theta=0$ and $\theta=\pi$. 

On the other hand, if the ansatz (\ref{ansatz}) is substituted into the field equations (\ref{pde}) for the cyclic coordinate $\phi^n$, the compatibility condition
\begin{equation}
    \omega^2\Gamma^n_{nn}-\Gamma^n_{ab}\dfrac{d \phi^a}{d x}\dfrac{d \phi^b}{d x}=0 \hspace{0.5cm},\hspace{0.5cm} \forall  a,b= 1,\dots ,n-1\label{condition0}\ ,
\end{equation}
has to be verified. This introduces some restrictions into the set of possible Sigma models having $Q$-solutions. In order for (\ref{condition0}) to hold, the following relations on the Christoffel symbols associated to the internal space can be imposed,
\begin{equation}
    \Gamma^{n}_{n n}=0 \hspace{0.5cm},\hspace{0.5cm} \Gamma^{n}_{ab}=0 \hspace{0.5cm},\hspace{0.5cm} \forall a,b=1,\dots,n-1 \ ,
\end{equation}
which imply that $g^{ij}$ has to be block-diagonal of the form: 
\begin{equation} \label{eq:metric_conditions}
    g^{ab}=g^{ab}(\widetilde{\phi}) \hspace{0.5cm},\hspace{0.5cm}  g^{a n}=0 \hspace{0.5cm},\hspace{0.5cm} g^{nn}=g^{nn}(\widetilde{\phi}) \hspace{0.5cm},\hspace{0.5cm} \forall  a,b= 1,\dots ,n-1 \ .
\end{equation}

Pluging together conditions (\ref{eq:metric_conditions}) and (\ref{ansatz}) into the field equations (\ref{pde}), finally the relations
\begin{equation}
    \dfrac{d^2\phi^a}{d x^2}+\Gamma^a_{bc}\dfrac{d \phi^b}{d x}\dfrac{d \phi^c}{d x}-\omega^2\Gamma^a_{n n}-g^{ab}\dfrac{\partial V}{\partial \phi^b}=0 \hspace{0.5cm},\hspace{0.5cm} \forall a,b,c =1,\dots,n-1 \label{pde2} \ ,
\end{equation}
fix the behavior of the static fields. Notice that (\ref{pde2}) is expressed as
\begin{equation}\label{pde3}
\dfrac{d^2\phi^a}{d x^2}+\Gamma^a_{bc}\dfrac{d \phi^b}{d x}\dfrac{d \phi^c}{d x}-g^{ab}\dfrac{\partial}{\partial\phi^b}\Big[V(\widetilde{\phi})-\dfrac{1}{2}\omega^2g_{nn}(\widetilde{\phi})\Big]=0
\end{equation}
and we introduce
\begin{equation}
 \overline{V}(\widetilde{\phi})= V(\widetilde{\phi})-\dfrac{1}{2}\omega^2 g_{nn}(\widetilde{\phi}) \ ,\hspace{0.5cm}\label{effectivepot}
\end{equation}
as the effective potential term. Equations (\ref{pde3}) can be interpreted as the equations of motion for a classical mechanical problem, where $x$ has to be understood as the mechanical time variable, and the mechanical potential is $-\overline{V}(\widetilde{\phi})$. Thus, the second order differential equations (\ref{pde2}) are nothing but the Euler-Lagrange equations for the static effective energy functional
\begin{equation}
\overline{E}[\widetilde{\phi}] = \int dx \Big[ \dfrac{1}{2} g_{ij} \dfrac{d\phi^i}{dx} \dfrac{d\phi^j}{dx} + \overline{V}(\widetilde{\phi}) \Big] \ , \label{ener03}
\end{equation}
and this fact involves the existence of a first integral in the form
\begin{equation}
I_1=\dfrac{1}{2} g_{ij} \dfrac{d\phi^i}{dx} \dfrac{d\phi^j}{dx} - \overline{V}(\widetilde{\phi}) \ .
\label{i1}
\end{equation}
Taking into account (\ref{eq:boundary}) and (\ref{eq:boundary2}), it is direct that necessarily $I_1$ vanishes, giving place to the Virial type relation
\begin{equation}
\dfrac{1}{2} g_{ij} \dfrac{d\phi^i}{dx} \dfrac{d\phi^j}{dx} = \overline{V}(\widetilde{\phi}) \ .\label{virial}
\end{equation}

\section{$Q$-ball/$Q$-kink type solutions in non-linear Sigma models}
In order to illustrate the theoretical framework discussed in Section 2, we shall analyze some non-linear Sigma models whose internal spaces are either the sphere $\mathbb{S}^2$, or the Horn Torus $\mathbb{T}^*$. In these cases the $Q$-ball/$Q$-kink solutions will be analytically identified.

\subsection{Non-linear $\mathbb{S}^2$-Sigma models with $Q$-balls/$Q$-kinks}
Let us consider a massive non-linear $\mathbb{S}^2$-Sigma model governed by the action functional
\begin{equation}
S = \int dx \int dt \Big[ \dfrac{1}{2} \, \partial_\mu \theta \, \partial^\mu \theta+\dfrac{1}{2} \, \sin^2\theta \, \partial_\mu \varphi \, \partial^\mu \varphi-V(\theta) \Big]
\end{equation}
where $(\theta,\varphi)$ are the usual spherical coordinates with $\theta \in [0,\pi]$ and $\varphi\in [0,2\pi)$ and the potential term $V(\theta)$ is given by
\begin{equation}\label{potential00}
    V(\theta)=\dfrac{1}{2}a^2\sin^2\theta = \dfrac{1}{4} a^2 (1-\cos (2\theta)) \ .
\end{equation}
Although the geometric formalism studied in this article is general for any target space, this case can be understood as the $\mathbb{S}^2$-sphere embedded in $\mathbb{R}^3$, where the scalar fields are constrained to satisfy, $\phi_1^2+\phi_2^2+\phi_3^2=1$ and thus $(\phi_1,\phi_2,\phi_3)$ are maps from the $(1+1)$-dimensional Minkowski space-time to a $\mathbb{S}^2$-sphere of radius 1, which is the target space of the model. The standard spherical coordinates $(\phi_1,\phi_2,\phi_3)$ in $\mathbb{S}^2$ are
\begin{equation*}
    \phi_1=\sin\theta\cos\varphi, \quad \phi_2=\sin\theta\sin\varphi, \quad  \phi_3=\cos\theta \ .
\end{equation*}

The expression (\ref{potential00}) corresponds to the sine-Gordon potential. In this case the system has two vacua located at the North and South poles of the sphere $\mathbb{S}^2$. This kind of structure has been studied previously in the context of static kinks in \cite{MAGLEON2008,MAGLEON2009}. In \cite{Loginov:2011zz} Loginov studied a $\mathbb{S}^2$-Sigma model with a sine-Gordon potential $V= \frac{1}{2} a^2 (1-\cos \theta)$, which in general has only one vacuum point located at the North Pole of $\mathbb{S}^2$. In this case, the generic $Q$-balls consisted of non-topological solitons asymptotically connecting this vacuum point to itself. As we will see, this fact represents a great difference with the model (\ref{potential00}), where the generic defects are topological $Q$-kinks connecting the two different vacua situated at the poles. In this case, the effective potential (\ref{effectivepot}) 
\begin{equation}
    \overline{V}(\theta)= \dfrac{1}{2}\left(a^2-\omega^2\right)\sin^2\theta
\end{equation}
adopts the same functional form as (\ref{potential00}), with a different coupling constant. The $Q$-kinks can now be analytically identified using relation (\ref{virial}), which leads to the expression
\begin{equation}\label{virial01}
    x-x_C = \displaystyle \int  \dfrac{d\theta}{\sqrt{\left(a^2-\omega^2\right)\sin^2\theta}} 
\end{equation}
where $x_C$ is an integration constant. From (\ref{virial01}) it is clear that the range of the internal rotation frequencies is:
\begin{equation}
   0 < \omega^2<a^2 \ .
\end{equation}

Integrating the equation (\ref{virial01}) and considering \ref{ansatz}, one obtains for the $Q$-kinks in spherical variables,
\begin{equation}\label{eq:KinkSolSphere}
    \begin{aligned}
    &\theta(x)=2 \arctan \left[ e^{\sqrt{a^2-\omega^2}\, (x-x_C)}\right] \ , \\
    &\varphi(t) = \omega t \ , 
    \end{aligned}
\end{equation}
where $x_C$ can be interpreted as the center of the solution. In Cartesian coordinates this solution reads:
\begin{equation}
    \begin{aligned}
    &\phi^{1}(t,x)=  \frac{\cos{\omega t}}{\cosh{\left(\sqrt{a^2-\omega^2}(x-x_{C})\right)}} \ , \\
    &\phi^{2}(t,x)=   \frac{\sin{\omega t}}{\cosh{\left(\sqrt{a^2-\omega^2}(x-x_{C})\right)}} \ , \\
    &\phi^{3}(t,x)=\tanh{\left(\sqrt{a^2-\omega^2}(x-x_{C})\right)} \ .
    \end{aligned}
\end{equation}

Plugging (\ref{eq:KinkSolSphere}) into the expressions of (\ref{energy01}) and (\ref{eq:Q_2dim}), it is direct to obtain the energy and the $Q$-charge for these solutions:

\begin{equation}
E\left(\omega\right)=\dfrac{2a^2}{\sqrt{a^2-\omega^2}} \hspace{0.5cm},\hspace{0.5cm} Q=\dfrac{2 \omega}{\sqrt{a^2-\omega^2}} \ ,
\end{equation}
where the energy can be written in terms of the charge as
\begin{equation}
    E\left(Q\right)=a\sqrt{Q^2+4} \ .
\end{equation}

\begin{figure}[H]
    \centering
    \includegraphics[width=75mm]{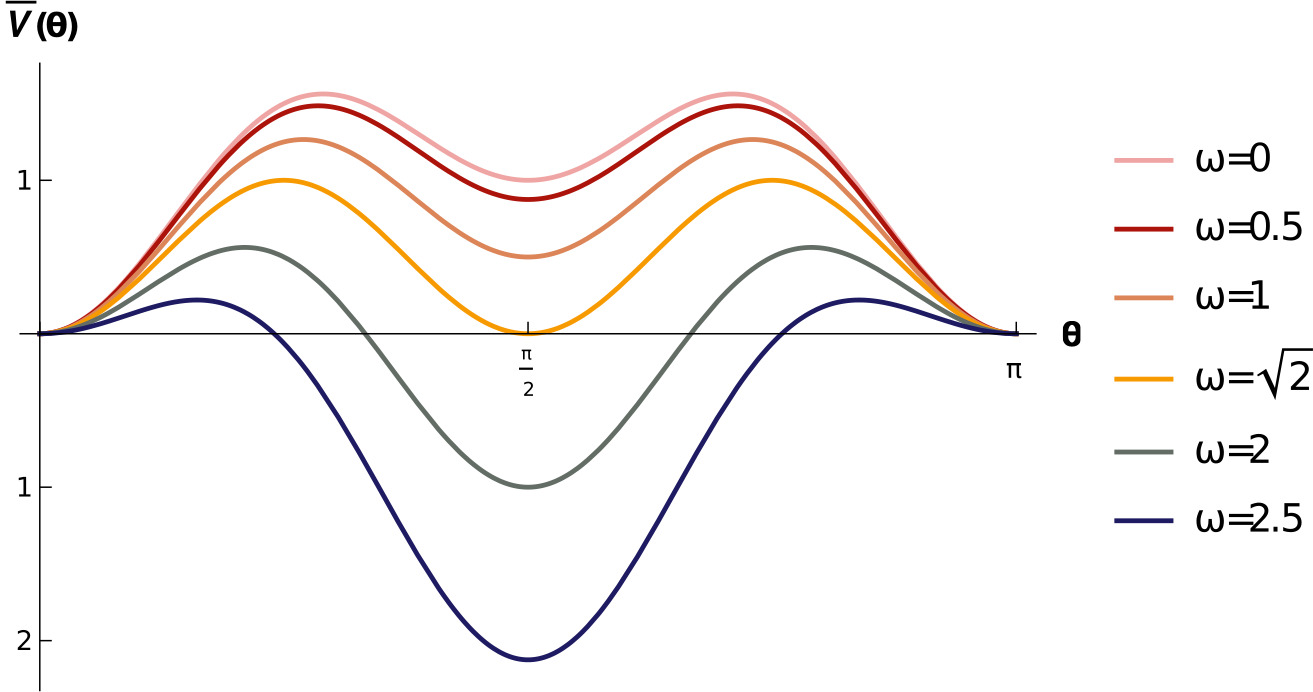}
    \caption{Effective potential $\overline V (\theta)$ in (\ref{eq:EffPotential2SG}) with $a=0.5$ for distinct values of the internal frequency $\omega$. } \label{fig:GraphEffPotential2SG}
\end{figure}

A second example, which introduces a novel behaviour is characterized by the potential\footnote{Indeed, the potential in (\ref{eq:DoubleSineGordonS2}) can be reduced to a more general expression of the form 
$ V(\theta) = \dfrac{1}{2}\left(A\sin^2\theta+B\sin^4\theta\right)$ by rescaling the coordinates and parameters for the system.}
\begin{equation}\label{eq:DoubleSineGordonS2}
    V(\theta)=1-(1-a)\cos 2\theta -a \cos (4\theta) = 2(1+3 a) \sin^2\theta - 8 a \sin^4 \theta \ .
\end{equation}
Formula (\ref{eq:DoubleSineGordonS2}) describes a double sine-Gordon potential, where for the particular values $a=0$ and $a=1$ the usual sine Gordon potential is recovered with different angular arguments. The potential must be positive semi-definite, which implies that $0\leq a \leq 1$ necessarily. The vacua are $\theta_{v}=0$ and $\theta_{v}=\pi$, and the effective potential (\ref{effectivepot}) is given by: 
\begin{equation}\label{eq:EffPotential2SG}
    \overline{V}\left(\theta\right)=\dfrac{1}{2}\sin^2\theta\left(4+12a-\omega^2 - 16 a \sin^2\theta\right)
\end{equation}
which allows the existence of $Q$-ball and/or $Q$-kink solutions depending on the value of $\omega$, as it can be seen in the Figure \ref{fig:GraphEffPotential2SG}. When the effective potential is also positive semi-definite, the existence of $Q$-kinks is guaranteed. On the other hand, when the internal frequency increases, a global minimum distinct from the vacua $\theta_{v}$ appears where $\overline V_{\rm{min}}<0$. In this case the conditions for the existence of $Q$-balls are met, since there exists a field value different from the vacuum at which the effective potential crosses linearly the axis $\theta$. From the previous argument, we shall distinguish two different regimes, which depend on the range of the internal rotational frequencies. In turn, the value of $a$ directly depends on the frequency range, where a larger value of $a$ results in a smaller lower bound for the internal frequency.

\begin{figure}[h] 
	\centering
    \subfigure{\includegraphics[width=65mm]{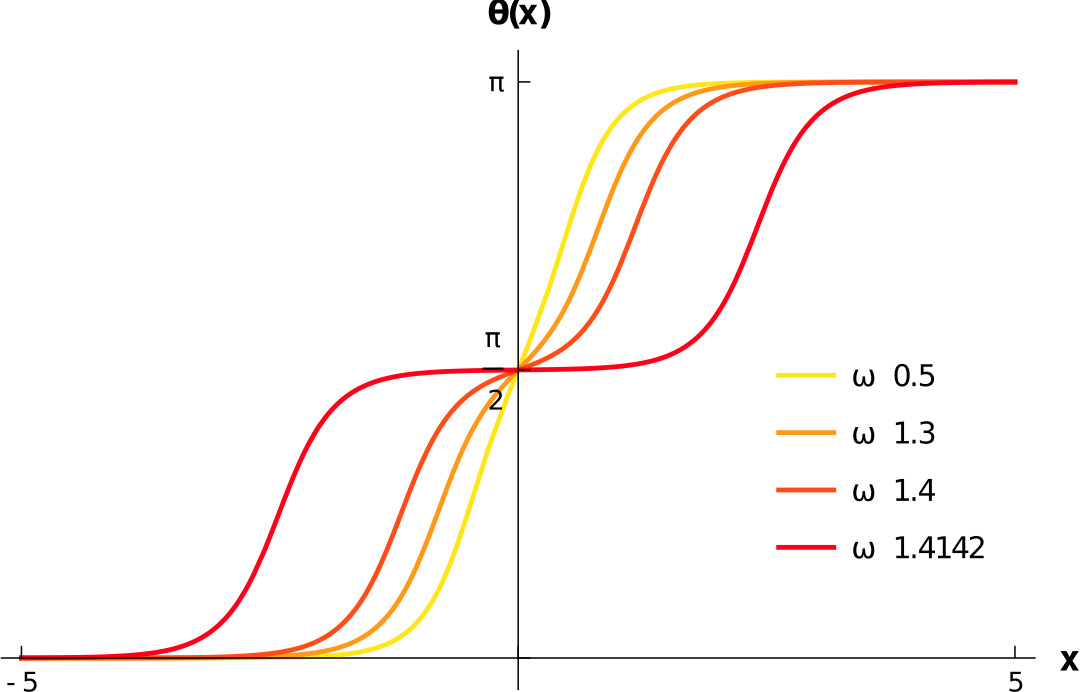}}
    \hspace{8mm}
    \subfigure{\includegraphics[width=35mm]{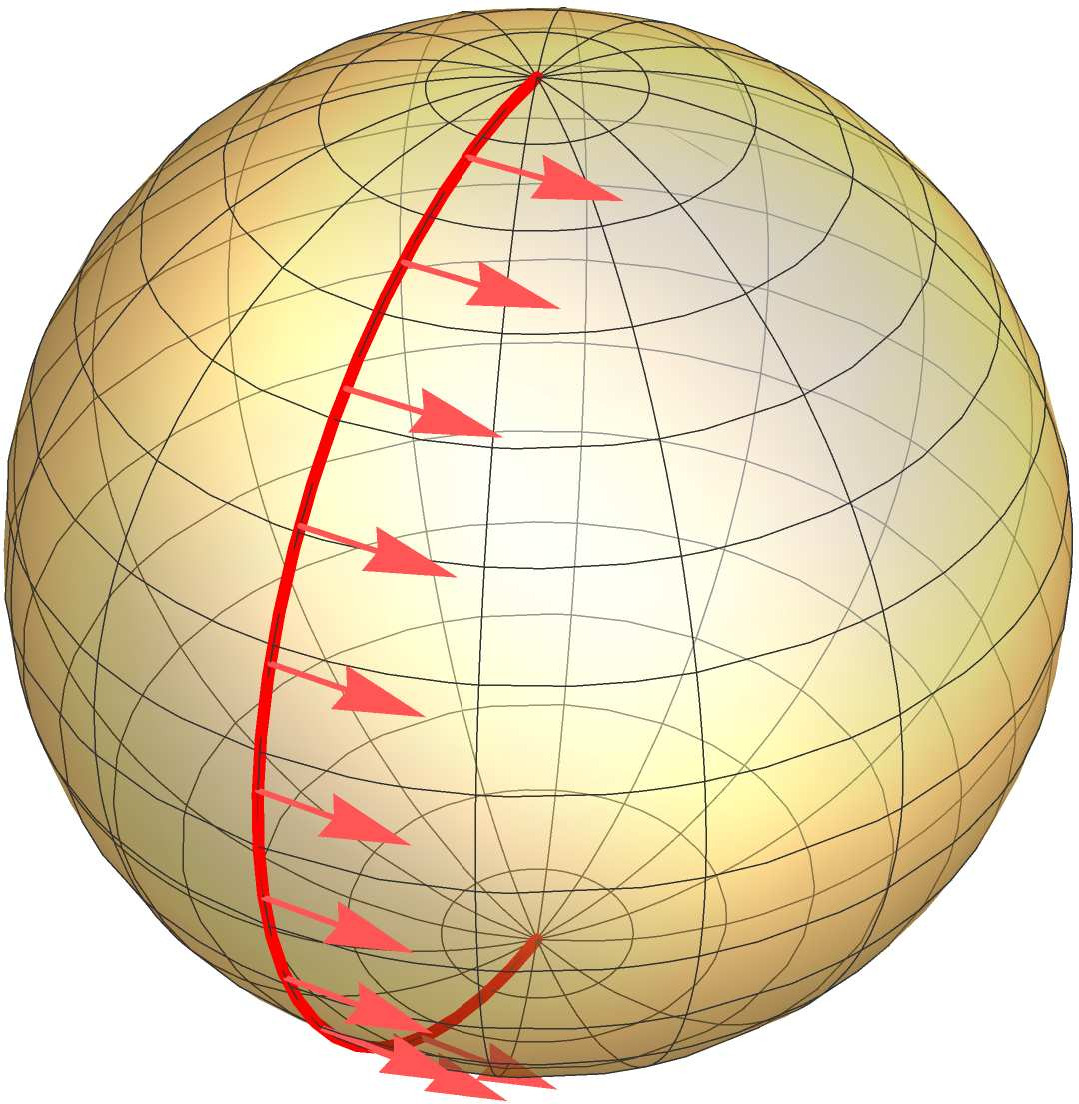}}
	\caption{Profiles (left) given by (\ref{eq:KinkSolSphere2}) with $a=0.5$ for different values of the internal frequency $\omega$ and orbit (right) for the $Q$-kinks in the Regime I.}
    \label{fig:QKinks_esfera}
\end{figure}

\paragraph{Regime I:} For the range $0 < \omega^2 < 4 (1-a)$ the effective potential is positive semi-definite. In this case the solutions derived from (\ref{virial}) consist of $Q$-kinks (see Figure \ref{fig:QKinks_esfera} left) whose expression is
\begin{equation}\label{eq:KinkSolSphere2}
    \begin{aligned}
    &\theta\left(x\right)=\arccos \left(\dfrac{\sqrt{-4+4 a+\omega ^2} \tanh \left(\sqrt{4+12 a-\omega ^2} \ (x-x_C) \right)}{\sqrt{-4+4 a+\omega ^2-16 \  a \ \text{sech}^2\left(\sqrt{4+12 a-\omega ^2} \ (x-x_C)\right)}}\right)\\
    &\varphi\left(t\right)=\omega t
    \end{aligned}
\end{equation}
which can be expressed in Cartesian coordinates as,
\begin{equation}
    \begin{aligned}
      &\phi^{1}(t,x)=  \sqrt{1+\frac{(-4+4a+\omega^2)\tanh^2{(\sqrt{4+12a-\omega^2}(x-x_{C}))}}{4-4a-\omega^2+16 a \text{ sech}^2(\sqrt{4+12a-\omega^2}(x-x_{C}))}} \cos{\omega t} \ ,  \\
      &  \phi^{2}(t,x)=  \sqrt{1+\frac{(-4+4a+\omega^2)\tanh^2{(\sqrt{4+12a-\omega^2}(x-x_{C}))}}{4-4a-\omega^2+16 a \text{ sech}^2(\sqrt{4+12a-\omega^2}(x-x_{C}))}} \sin{\omega t} \ ,   \\
      & \phi^{3}(t,x)=\frac{\sqrt{-4+4a+\omega^2}\tanh({\sqrt{4+12a-\omega^2}(x-x_{C})})}{\sqrt{4-4a-\omega^2+16 a \text{ sech}^2(\sqrt{4+12a-\omega^2}(x-x_{C}))}} \ .
    \end{aligned}
\end{equation}

This solution interpolates between two different vacua, the North and South poles (the corresponding behaviour on $\mathbb{S}^2$ is represented in Figure \ref{fig:QKinks_esfera} right), and for this reason it is not possible to continuously deform it to one of the vacua with finite energy. Therefore, $Q$-kinks and the vacuum solutions belong to different topological sectors. 

The energy and the $Q$-charge for these solutions are,
\begin{equation}
\begin{aligned} \label{eq:Energy_and_charge_Reg1_sphere}
    &E\left(\omega\right)=\sqrt{4+12 a-\omega ^2}+\dfrac{\left(4-4 a+\omega ^2\right) \text{arctanh}\left(\dfrac{4 \sqrt{a}}{\sqrt{4+12 a-\omega ^2}}\right)}{4 \sqrt{a}} \ ,\\
    &Q(\omega)=\dfrac{\omega}{2 \sqrt{a}} \ \text{arctanh}\left(\dfrac{4 \sqrt{a}}{\sqrt{4+12 a-\omega ^2}}\right) \ .
\end{aligned}
\end{equation}

\begin{figure}[h] 
	\centering
    \subfigure{\includegraphics[width=65mm]{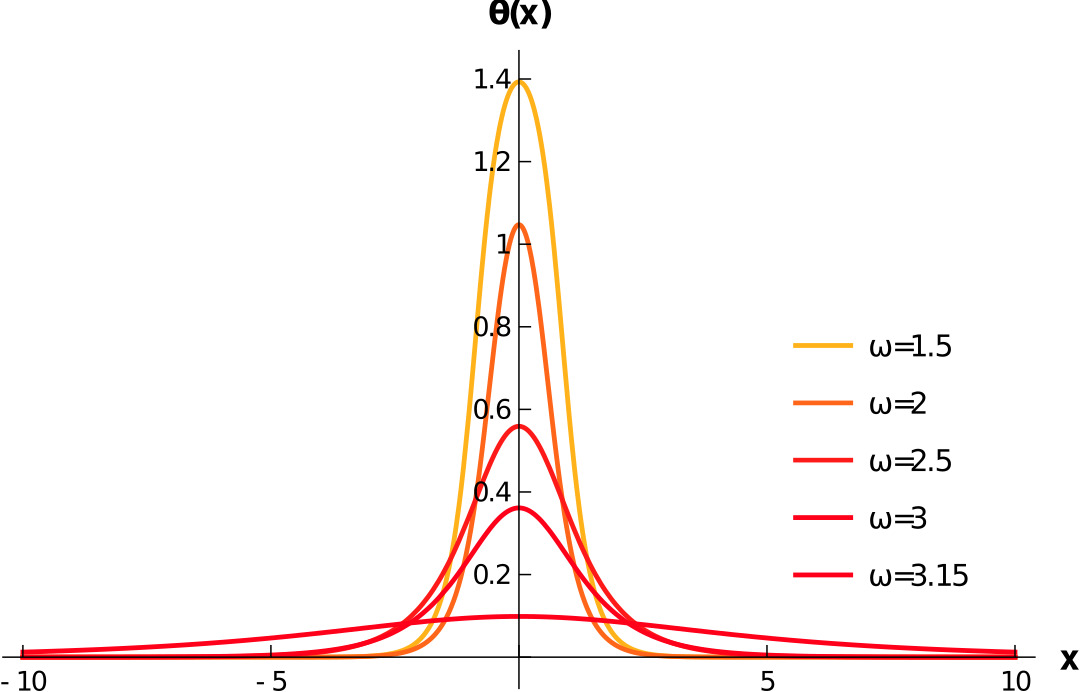}}
    \hspace{8mm}
    \subfigure{\includegraphics[width=35mm]{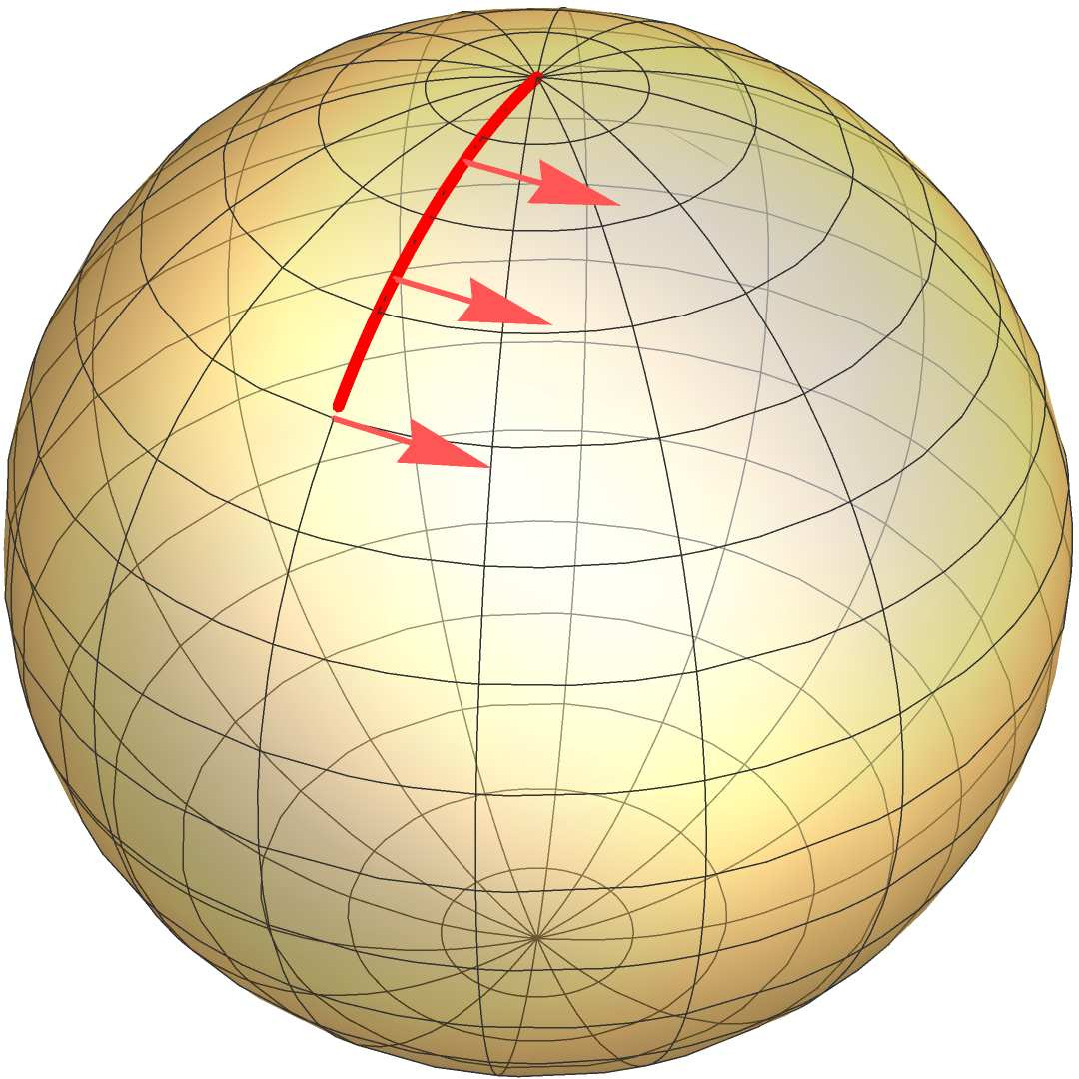}}
	\caption{Profiles (left) given by (\ref{eq:QBallSolSphere2}) with $a=0.5$ for different values of the internal frequency $\omega$ and orbit (right) for $Q$-balls in the Regime II.}
    \label{fig:QBalls_esfera}
\end{figure}

\paragraph{Regime II:} For the range $4(1-a)<\omega^2<4(1+3a)$, the effective potential acquires a single global minimum such that $\overline V_{\rm{min}}<0$. In this scenario, the solutions coming from (\ref{virial}) are $Q$-balls (see Figure \ref{fig:QBalls_esfera} left). The analytic expression of the field $\theta(x)$ is:
\begin{equation}\label{eq:QBallSolSphere2}
    \theta\left(x\right)=\arccos\left(-\dfrac{\sqrt{4-4 a-\omega ^2}}{\sqrt{\left(4+12 a-\omega ^2\right) \tanh ^2\left(\sqrt{4+12 a-\omega ^2}(x-x_C)\right)-16 a}}\right) \ ,
\end{equation}
which implies a time depending solutions in Cartesian coordinates as,
\begin{equation}
    \begin{aligned}
      &\phi^{1}(t,x)=  \sqrt{1+\frac{-4+4a+\omega^2}{(4+12a-\omega^2)\tanh^2{(\sqrt{4+12a-\omega^2}(x-x_{C}))}-16 a) }} \cos{\omega t} \ , \\
      &  \phi^{2}(t,x)=  \sqrt{1+\frac{-4+4a+\omega^2}{(4+12a-\omega^2)\tanh^2{(\sqrt{4+12a-\omega^2}(x-x_{C}))}-16 a) }} \sin{\omega t} \ ,\\
      & \phi^{3}(t,x)= -\frac{\sqrt{4-4a-\omega^2}}{\sqrt{(4+12a-\omega^2)\tanh^2{(\sqrt{4+12a-\omega^2}(x-x_{C}))}-16 a)}} \ ,
    \end{aligned}
\end{equation}
in Cartesian coordinates. These solutions connect a vacuum with itself and in principle, it could be possible to continuously deform them with finite energy to the vacuum. Therefore, this solution belongs to the same topological sector that the vacuum, and its topological charge is zero. 

The energy and the conserved charge are now, 
\begin{equation}
    \begin{aligned} \label{eq:Energy_and_charge_Reg2_sphere}
     &E(\omega) = \sqrt{4+12 a-\omega ^2}+\dfrac{\left(4-4 a+\omega ^2\right) \text{arccoth}\left(\dfrac{4 \sqrt{a}}{\sqrt{4+12 a-\omega ^2}}\right)}{4 \sqrt{a}} \ , \\
     &Q(\omega)=\dfrac{\omega}{2 \sqrt{a}}\text{arccoth}\left(\dfrac{4 \sqrt{a}}{\sqrt{4+12 a-\omega ^2}}\right) \ .
     \end{aligned}
\end{equation}

\begin{figure}[h] 
	\centering
    \subfigure{\includegraphics[width=65mm]{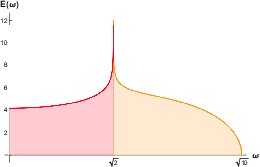}}
    \hspace{8mm}
    \subfigure{\includegraphics[width=65mm]{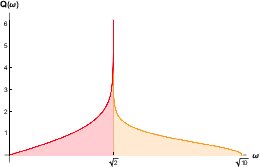}}
	\caption{Energy (left) and $Q$-charge (right) of the $Q$-kink (\ref{eq:Energy_and_charge_Reg1_sphere}) (red) and $Q$-ball (\ref{eq:Energy_and_charge_Reg2_sphere}) (orange) as a function of the frequency $\omega$ for the value of the parameter $a=0.5$.}
    \label{fig:Energia_Carga_esfera}
\end{figure}

Between these two Regimes, in the case  $\omega^2 = 4(1-a)$, the effective potential has three absolute minima, the two original vacua and another one in $\theta=\pi/2$. In this scenario, both energy and Noether charge are not finite (see Figure \ref{fig:Energia_Carga_esfera}). It means that the transition between the regimes is impossible because it would require infinite energy. This is an interesting behaviour and indicates that $Q$-balls and $Q$-kinks belong to different topological sectors, as we have previously mentioned. Therefore, transition between both sectors is not allowed.

\subsection{Non-linear $\mathbb{T}^*$-Sigma model with $Q$-balls/$Q$-kinks}
A non-linear $\mathbb{T}^*$-Sigma model is now considered, where the internal space is the \textit{Horn Torus} $\mathbb{T}^*$. The Horn Torus can be defined as the torus where the distance from the center of the tube to the center of the torus is the same as the radius of the tube. Thus, let consider a Horn Torus embedded in $\mathbb{R}^3$ and which in Cartesian coordinates reads,
\begin{equation}
  \begin{aligned}
    &\phi^1= \cos{\varphi}\left(1+\cos{\theta}\right) \ ,  \\
    &\phi^2= \sin{\varphi}\left(1+\cos{\theta}\right) \ , \\
    &\phi^3=\sin{\theta},
  \end{aligned}
\end{equation}
where $(\theta,\varphi)$ are toroidal-poloidal coordinates with $\theta \in [-\pi,\pi)$ and $\varphi\in [0,2\pi)$. In this context we shall study two different models. The first one involves the action functional,
\begin{equation}
	S = \int dx \int dt \left[\dfrac{1}{2}\partial_\mu \theta \partial^\mu \theta+\dfrac{1}{2}(1+\cos\theta)^2\partial_\mu \varphi \partial^\mu \varphi-V(\theta)\right]
\end{equation}
where the potential term has been chosen as
\begin{equation} \label{eq:potential_torus}
	V(\theta)=\dfrac{1}{2}(1+\cos\theta)^2 \ .
\end{equation}
The vacuum of the theory is located at $\theta_{v}=-\pi$ which corresponds to the centre of the Horn Torus. The effective potential is,
\begin{equation} \label{eq:potential_eff_torus}
	\overline{V}(\theta)=\dfrac{1}{2}(1-\omega^2)(1+\cos\theta)^2 \ .
\end{equation}

In this case, the internal frequencies are restricted to the following values
\begin{equation}
  0<\omega^2<1 \ ,
\end{equation}
for which the $\theta$-field of the $Q$-kink reads:
\begin{equation}\label{eq:Profile_Torus}
	\theta(x)=2 \arctan\left( \sqrt{1-\omega ^2}(x-x_C)\right) \ .
\end{equation} 

\begin{figure}[H]
\centering
\includegraphics[width=45mm]{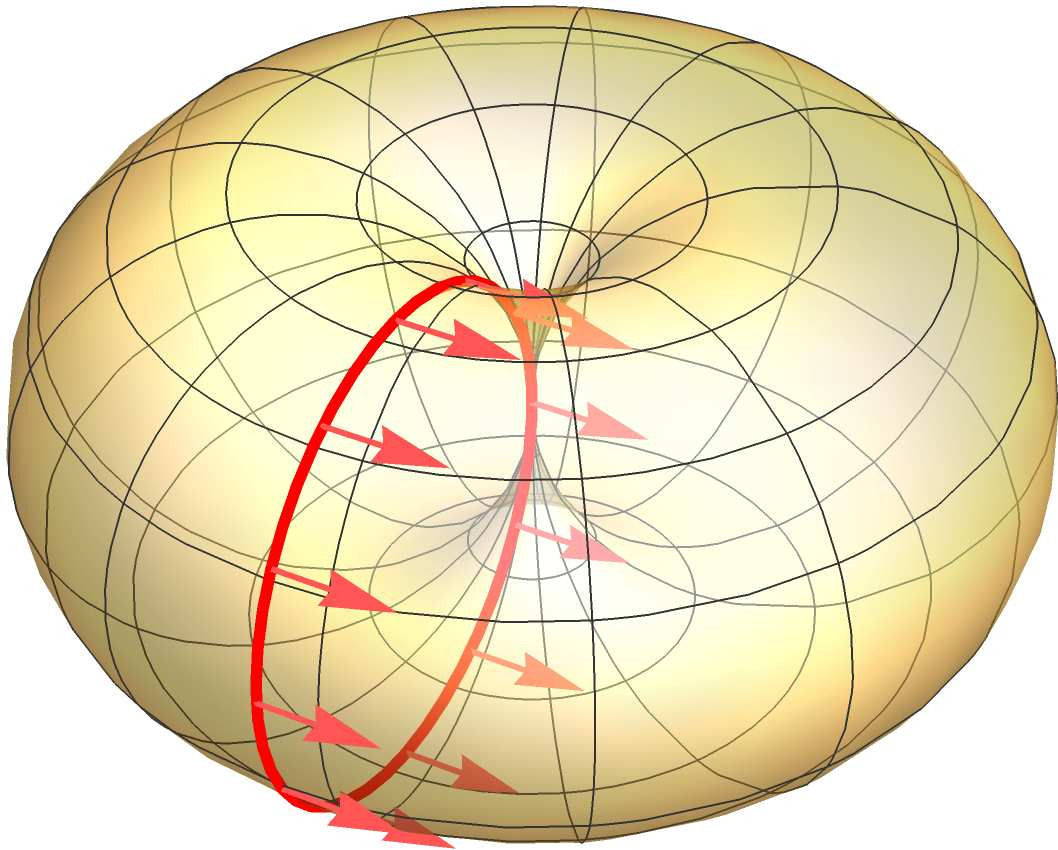}
\caption{$Q$-kink orbit on the surface of the Horn Torus spinning with an internal frequency $\omega$.} \label{fig:TorusKinks}
\end{figure}

The only existing solutions are $Q$-kinks. Since these objects connect the vacuum with itself, they can be understood as non-topological kinks that rotate with an internal rotational frequency around the tube of the torus as it is illustrated in the Figure \ref{fig:TorusKinks}. These solutions are spinning ``vrochosons" and belong to a different topological sector than the vacuum due to the topology of the target space. The term vrochoson was introduced in \cite{Alonso2023_toro_1}, referred to non-topological kinks that are globally stable because of the topological properties of the target space, as it is the case. More examples of vrochosons can be found in \cite{Alonso2023_toro_1, Alonso2022_toro_2,Alonso2007_toro3}, where different nonlinear $(\mathbb{S}^1\times \mathbb{S}^1)$-Sigma models are studied. The $Q$-kinks written in Cartesian coordinates are given by
\begin{equation}
    \begin{aligned}
    &\phi^{1}(t,x)= \dfrac{2 \cos (\omega t )}{1-\left(\omega ^2-1\right) (x-x_C)^2} \ ,\\
    &\phi^{2}(t,x)= \dfrac{2 \sin (\omega t)}{1-\left(\omega ^2-1\right) (x-x_C)^2} \ ,\\
    &\phi^{3}(t,x)= \dfrac{2 \sqrt{1-\omega ^2}(x-x_C)}{1-\left(\omega ^2-1\right) (x-x_C)^2} \ .\\
    \end{aligned}
\end{equation}

Finally, the energy and the Noether charge associated to (\ref{eq:Profile_Torus}) are
\begin{equation}
    E(\omega)=\dfrac{2\pi}{\sqrt{1-\omega^2}} \ , \quad
    Q(\omega)=\dfrac{2\pi \omega}{\sqrt{1-\omega^2}} \ ,
\end{equation}
which allow to write one in terms of the other as:
\begin{equation}
    E(Q)=\sqrt{4\pi^2+Q^2} \ .
\end{equation}

The behaviour of the energy and the $Q$-charge as a function of the frequency $\omega$ is completely similar to the first $\mathbb{S}^2$-Sigma model exposed in the previous subsection. 

\begin{figure}[H]
    \centering
    \includegraphics[width=75mm]{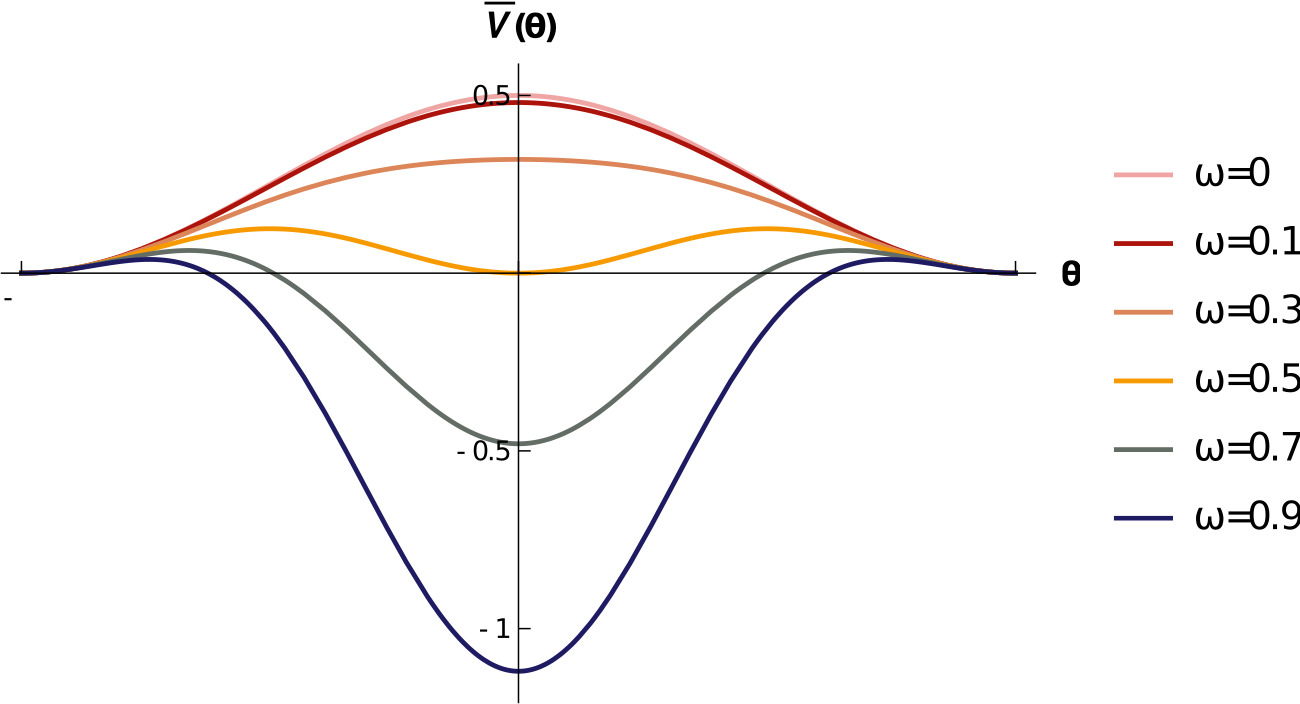}
    \caption{Effective potential $\overline V (\theta)$ given in  (\ref{eq:EffPotentialTorusQB-QK}) for distinct values of the internal frequency $\omega$. } \label{fig:EffPotential2SG}
\end{figure}

As we have seen in the case of the sphere, it is possible to find Sigma models in which $Q$-ball and $Q$-kink solutions coexist. Let us consider the potential
\begin{equation}
    V=\dfrac{1}{4}\left(1+\cos\theta\right)
\end{equation}
where the vacuum of the potential is $\theta_v=-\pi$. The effective potential corresponds to the following formula,
\begin{equation}\label{eq:EffPotentialTorusQB-QK}
    \overline{V}(\theta)=\dfrac{1}{4} (1+\cos\theta) \left(1-2 \omega ^2 (1+\cos\theta)\right) \ .
\end{equation}
The form of the potential in (\ref{eq:EffPotentialTorusQB-QK}) guarantees the existence of solutions of $Q$-ball and $Q$-kink type depending on the value of the frequency $\omega$. The shape of the potential for different values of the frequency $\omega$ are shown in Figure \ref{fig:EffPotential2SG}. For $\omega<1/4$, the only global minimum of the effective potential is at the vacuum. Therefore, the effective potential is positive semi-definite and the only valid solutions are of $Q$-kink type. For $\omega>1/4$, a new global minimum distinct from the vaccum appears at $\theta=0$ where $\overline{V}_{\text{min}}<0$. Hence, in this region the existence of $Q$-balls is guaranteed. From distinct values of the internal rotational frequencies, two different regimes can be distinguished.  Each of these two sectors will be examined in detail below.

\begin{figure}[h] 
	\centering
    \subfigure{\includegraphics[width=65mm]{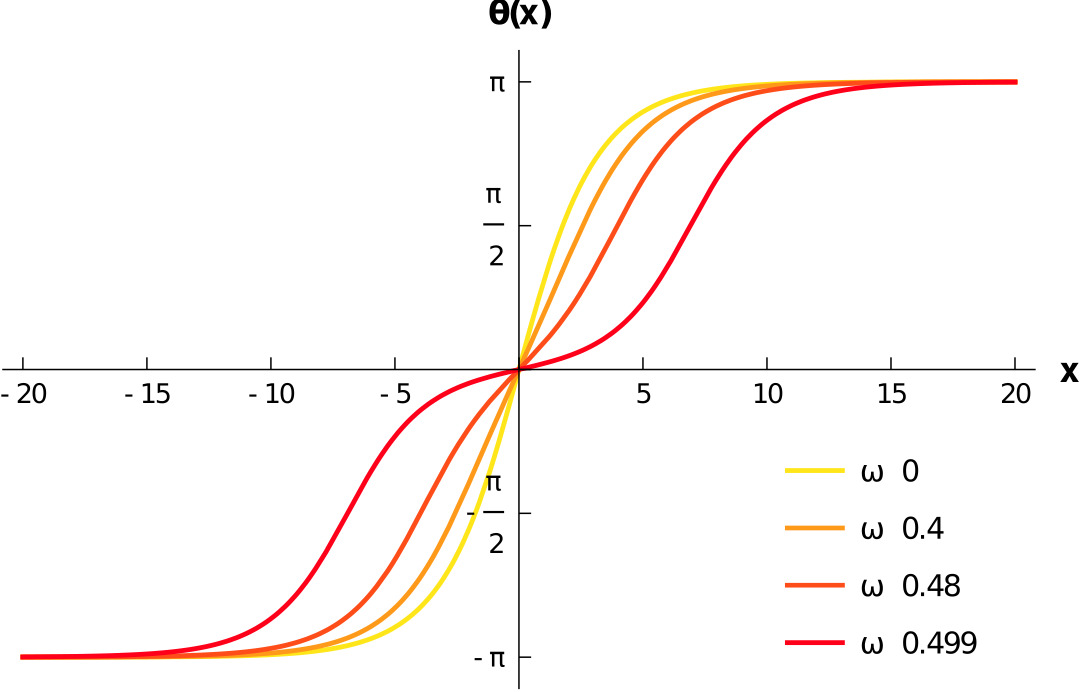}}
    \hspace{8mm}
    \subfigure{\includegraphics[width=35mm]{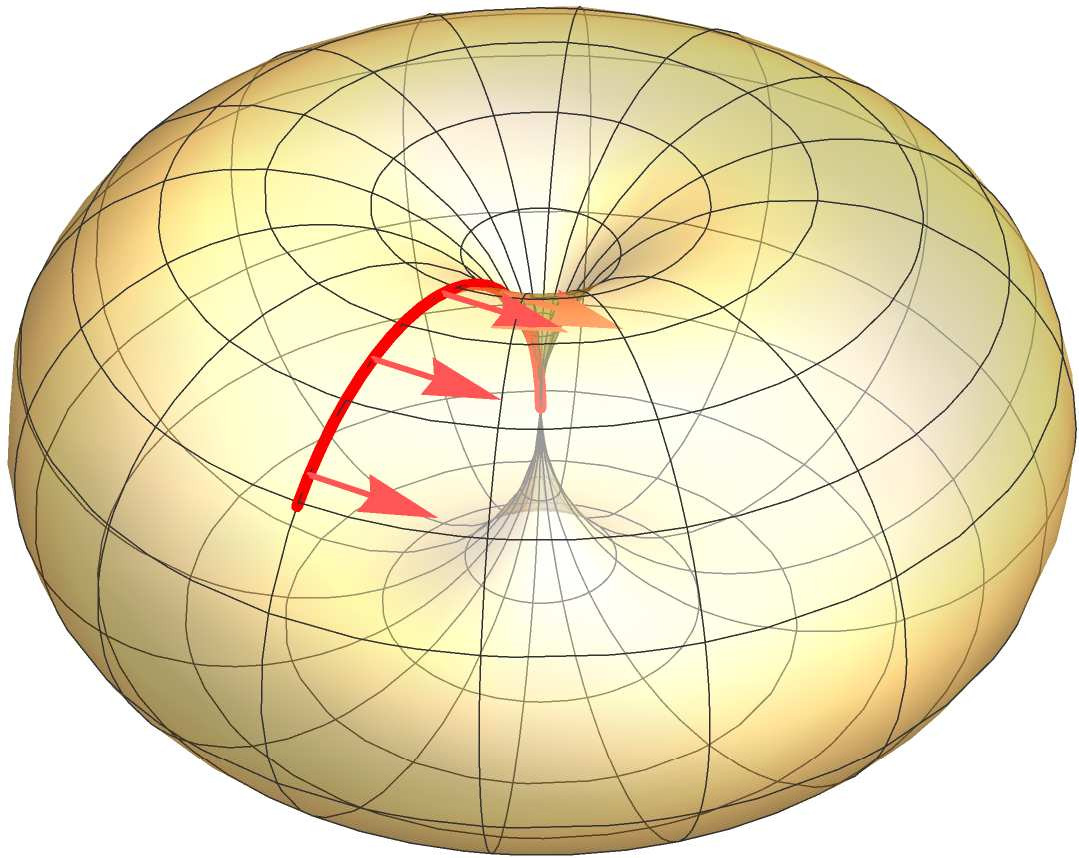}}
	\caption{Profiles for different values of the internal frequency $\omega$ (left) and orbit on the torus (right) of the $Q$-kink solutions (Regime I).}
    \label{fig:QKinks_toro}
\end{figure}

\paragraph{Regime I:} For the range $0 < \omega^2 < \frac{1}{4}$ we obtain the $\theta$ field  for the $Q$-kink type solutions (see Figure \ref{fig:QKinks_toro})
\begin{equation}
    \theta(x)=2 \text{arcsin}\left(\dfrac{\sqrt{1-4 \omega ^2} \tanh \left(\frac{x-x_{C}}{2}\right)}{\sqrt{1-4 \omega ^2 \tanh ^2\left(\frac{x-x_{C}}{2}\right)}}\right)
\end{equation}
which in Cartesian coordinates are
\begin{equation}
    \begin{aligned}
        &\phi^{1}(t,x)= \dfrac{4 \cos \omega t}{1+4 \omega^2+(1-4\omega^2)\cosh (x-x_{C}))} \ ,\\
        &\phi^{2}(t,x)= \dfrac{4 \sin \omega t}{1+4 \omega^2+(1-4\omega^2)\cosh  (x-x_{C})} \ ,\\
        &\phi^{3}(t,x)= \dfrac{4 \sqrt{1-4\omega^2} \sinh{(\frac{ x-x_{C}}{2})}}{1+4 \omega^2+(1-4\omega^2)\cosh  (x-x_{C})} \ .
\end{aligned}
\end{equation}
This type of solutions have a similar qualitative behaviour as the previously model for $Q$-kinks in the $\mathbb{S}^2$-sphere. As we can see in Figure \ref{fig:QKinks_toro}, when the frequency increases, a plateau appears in the profile of the kink. The reason is that the effective potential at the local minimum $\theta=0$, tends to zero when $\omega^2 = 4(a-1)$.  

The energy and Noether charge are,
\begin{equation}
    \begin{aligned}\label{eq:Energy_and_charge_Reg1_toro}
    &E(\omega)=\dfrac{2}{\omega}\text{arctanh}(2 \omega), \quad Q(\omega)=\dfrac{1}{\omega^2}\left[(1+4\omega^2)\text{arctanh}(2 \omega)-2\omega\right] \ .
    \end{aligned}
\end{equation}

\paragraph{Regime II:} For the range $\omega^2>\frac{1}{4}$, the profile of static field $\theta$ of the $Q$-ball type solutions is:
\begin{equation}
    \theta(x)=2 \ \text{arccsc}\left(\sqrt{\dfrac{\tanh^2\left(\frac{x-x_{C}}{2}\right)-4 \omega ^2}{1-4 \omega ^2}}\right)  \ .
\end{equation}

The $Q$-ball in Cartesian coordinates is:
\begin{equation}
    \begin{aligned}
        &\phi^{1}(t,x)= \dfrac{4 \cos \omega t}{1+4 \omega^2+(4\omega^2-1)\cosh(x-x_{C})} \ ,\\
        &\phi^{2}(t,x)= \dfrac{4 \sin \omega t}{1+4 \omega^2+(4\omega^2-1)\cosh(x-x_{C})} \ ,\\
        &\phi^{3}(t,x)= \frac{2 \sqrt{4 \omega^2-1} \ \text{sech}({\frac{x-x_{C}}{2}})}{4 \omega^2-\tanh^2{(\frac{x-x_{C}}{2})}} \ .
    \end{aligned}
\end{equation}
However, in contrast to what happens in most of the models where $Q$-balls exist, there is not an upper bound of the internal frequency $\omega$. The reason is that $ \left. \frac{d^2 \overline{V}(\theta)}{d \theta^2}\right|_{\theta=\theta_{v}}>0$ for all $\omega$. The energy and Noether charge obtained are
\begin{equation}
    \begin{aligned}\label{eq:Energy_and_charge_Reg2_toro}
    &E=\dfrac{2}{\omega}\text{arctanh}\left(\dfrac{1}{2 \omega}\right), \quad  Q=\dfrac{1}{\omega ^2}\left[\left(1+4 \omega ^2\right) \text{arccoth}(2 \omega )-2 \omega \right] \ .
    \end{aligned}
\end{equation}

The vacuum and the solution of $Q$-ball type belong to the same topological sector. This can be easily understood since such solutions can be continuously deformed to the vacuum with finite energy.

\begin{figure}[h] 
	\centering
    \subfigure{\includegraphics[width=65mm]{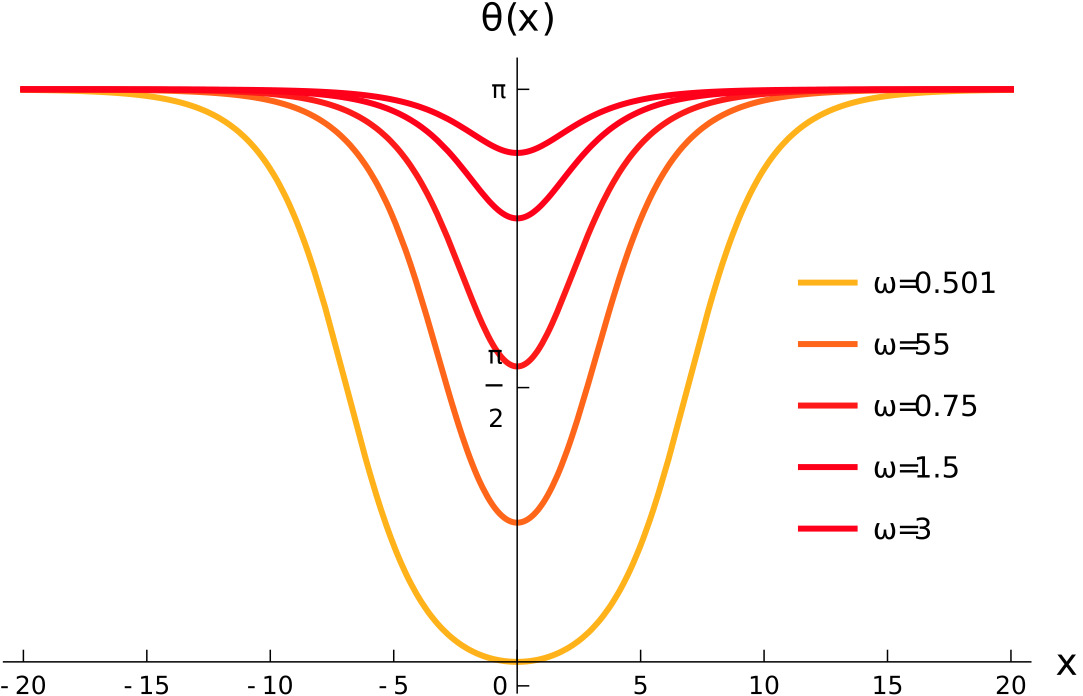}}
    \hspace{8mm}
    \subfigure{\includegraphics[width=35mm]{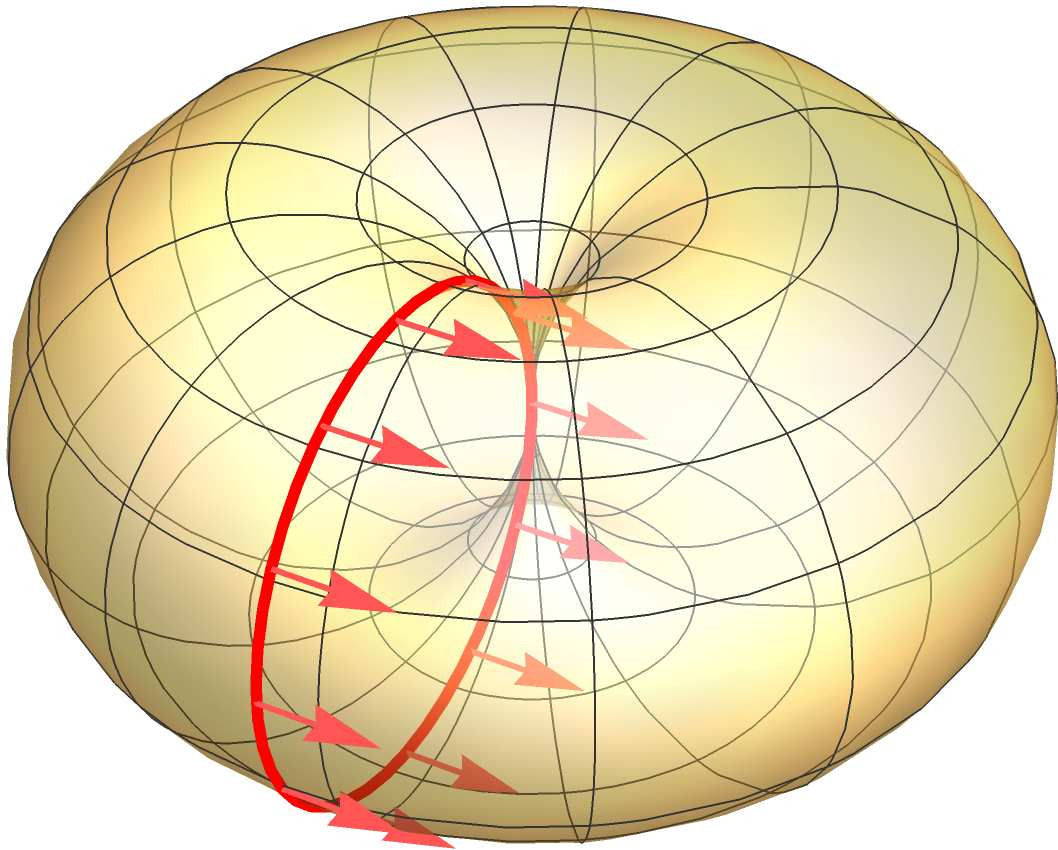}}
	\caption{Profiles for different values of the internal frequency $\omega$ (left) and orbit on the torus (right) of the $Q$-balls solutions (Regime II).}
 \label{fig:$Q$-balls_toro}
\end{figure}

\begin{figure}[H] 
	\centering
    \subfigure{\includegraphics[width=65mm]{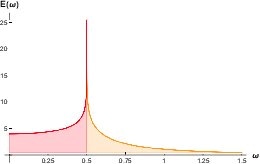}}
    \hspace{8mm}
    \subfigure{\includegraphics[width=65mm]{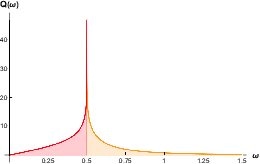}}
	\caption{Energy (left) and $Q$-charge (right) of the $Q$-kink (\ref{eq:Energy_and_charge_Reg1_toro}) (red) and $Q$-ball (\ref{eq:Energy_and_charge_Reg2_toro}).}
    \label{fig:Energia_Carga_toro}
\end{figure}

For $\omega^2=1/4$, at the transition between the two regimes takes place. The two regions are not connected, since the energy and the $Q$-charge required for one solution to go from one sector to the other are infinity (see Figure \ref{fig:Energia_Carga_toro}). Hence, there are two topologically disconnected regions as consequence of the Horn Torus topology.

\section{Conclusions and further comments}
This paper develops a geometric formalism for non-linear Sigma models which allows us to understand $Q$-ball and $Q$-kink solutions under the same perspective. This theoretical framework exploits the existence of a cyclic coordinate that leads to the invariance of the functional action under $U(1)$ symmetry group transformations. This, together with the potential term and the adequate conditions of the metric tensor, let us to find non-linear Sigma models where explicit analytical solution of $Q$-ball and/or $Q$-kink type are identified, and interesting properties arise.

In order to illustrate the theoretical results, we have applied this scheme to different non-linear Sigma models with distinct target spaces. In particular, we have studied models where the internal spaces are the $\mathbb{S}^2$-sphere and the Horn Torus, and it is possible to find analytical solutions for $Q$-balls/$Q$-kinks. In fact, we have found examples where both type of solutions appear for different frequencies $\omega$. For this situation, there exist two disconnected sectors, one with $Q$-kink solutions and the other with $Q$-ball solutions. At the limit between of these two regimes, both the energy and Noether charge tend to infinity, so it is not possible the transition between the solution of these two sectors. In the case of the non-linear $\mathbb{S}^2$-Sigma model, the $Q$-kinks connect two distinct vacua, whereas for $\mathbb{T}^*$-Sigma model, the $Q$-kinks connect a vacuum point with itself in a non-contractible way. These solutions can be understood as spinning vrochosons, which are non-topological kinks where the topology of the target space prevents them from decaying into vacuum.

The proposed geometrical formalism allows us to find either $Q$-balls or $Q$-kinks in certain Sigma models with one Noether charge. However, the theoretical framework is not restricted to just one conserved charge, it is possible to consider multiple cyclic variables. Therefore, the formalism introduced in the present paper opens up possibilities for future works where systems with multiple internal frequencies of $Q$-balls and $Q$-kinks could be present.  

\section*{Acknowledgements}
This research was funded by the Spanish Ministerio de Ciencia e Innovaci\'on (MCIN) with funding from the European Union NextGenerationEU (PRTRC17.I1) and the Consejer\'{\i}a de Educaci\'on, Junta de Castilla y Le\'on, through QCAYLE project, as well as grant PID2020-113406GB-I00 MTM  funded by MCIN/AEI/10.13039/501100011033.

\bibliographystyle{unsrt}

\end{document}